\documentclass[preprint,
,amssymb
,nofootinbib
]{revtex4-1}
\usepackage{epstopdf}
\usepackage{graphicx}
\usepackage{subfigure}
\usepackage{color}
\usepackage{amsmath}
\usepackage{epsfig,graphicx,bm}
\usepackage{epstopdf, epsf}
\usepackage{epstopdf}
\usepackage{dcolumn}
\usepackage{hyperref}
\everymath{\displaystyle}
\begin{document}
\title{Condensation of a scalar field non-minimally coupled to gravity in a cosmological context}  
\author{Amir Ghalee} 
\affiliation{{Department of Physics, Tafresh  University,
P. O. Box 39518-79611, Tafresh, Iran}}
\begin{abstract}
We present a new mechanism to condense a scalar field coupled to the Gauss-Bonnet term. We propose a scenario in which the condensed state will emerge from the background energy density in the late-Universe. During the radiation and dust-dominated eras, the energy density of the scalar field, $\rho_{\varphi}$, decreases at a slower rate than the background density. Eventually, $\rho_{\varphi}$ dominates over the energy density of dust and the scalar field could be condensed. In the condensed phase, we have the de Sitter phase for the universe  with $\rho_{\varphi}=3H_0^2M_P^2$. Moreover, we study the cosmological perturbations of the model and explore predictions of the model.
\end{abstract}
\maketitle
\section{INTRODUCTION}
Observations of the cosmic microwave background and large scale structure with the luminosity distances of high redshift supernova show that the universe is presently undergoing accelerated expansion \citep{{per},{rie},plank-data}.\\
In the $\Lambda$CDM model, the existence of an accelerated expansion era is explained  by adding the cosmological constant term , $\mathcal{L}_\Lambda=-\sqrt{-g}M^2_P\Lambda $, where $M_{P}^{2}$ is the reduced Planck mass, to the Einstein-Hilbert (EH) action. The cosmological constant term produces an energy density, $\rho_{\Lambda}=\Lambda M_P^2$, which is related to the present value of the Hubble parameter, $H_0$, as $\rho_{\Lambda}=3M_P^2H_0^2\approx10^{-3}eV^4$.\\
Although the fate of the universe is determined by $\rho_{\Lambda}$, but it seems very difficult to explain it by fundamental arguments (models) \citep{{wein},{jad},{vafa}}. So, various attempts have been made to explain the observations by  phenomenological models. For example, in Ref. \citep{Arkani} Arkani-Hamed et al. propose the ghost condensate scenario as
\begin{equation}
 \mathcal{L}=\sqrt{-g}M^4P(X), \hspace{1cm} X\equiv\frac{-\partial_\mu \varphi\partial^\mu\varphi}{2M^4},
\end{equation}
where $M$ is a scale of energy, to explain the current phase of the universe. It has been shown that if $P(X)$ has a minimum at $X=X_m$, with $P(X_m)\neq0$, the model leads to an accelerated expansion of the universe \citep{Arkani}. To suppress dynamics of $X$ during the radiation and dust-dominated eras, $X$ must be close to the minimum of $P(X)$ .\\
On the other hand, the successes of the standard model of cosmology show that any additional energy density should be subdominant during the radiation and dust-dominated eras. This issue can be solved by tracker fields \citep{{traka},{trakb}}. In such models, during the radiation and dust-dominated eras, the corresponding energy density in the tracker field decreases at a slower rate than the energy density in the background. Eventually, the corresponding energy density of the tracker field dominates over the dust density in such a way that leads to an accelerating universe \citep{{traka},{trakb}}.\\
Reconciling the above statements with a dynamical mechanism is not an easy task. A notable example is the quintessence model, which is described by the following Lagrangian
\begin{equation}
\mathcal{L}=-\frac{1}{2}\partial_{\mu}\phi\partial^{\mu}\phi-\frac{M^{4+\alpha}}{\phi^{\alpha}}, 
\end{equation}
where $\alpha$ is an arbitrary positive constant  \citep{{traka},{trakb},{trakc}}.\\
Although higher derivative theories of gravity have been used to study the primordial inflation but, by using suitable form, they can be used for current phase of the universe \citep{{modi},{modr}}. For instance, the Gauss-Bonnet term which is defined by
\begin{equation}
R_{GB}^2\equiv R^2-4R^{\alpha\beta}R_{\alpha\beta}+R^{\alpha\beta\gamma\eta}R_{\alpha\beta\gamma\eta}\hspace{0.3cm},
\end{equation}
has been used in some cosmological models. The fact that $\int\sqrt{-g}R_{GB}^2d^4x$ is a topological invariant, provides motivations to construct some phenomenological models \citep{{Gbr},{Gbr2},{Gbr3}}.\\
In this paper, we consider the following model 
\begin{equation}\label{i-0}
S=\int d^{4}x\sqrt{-g}\left[\frac{M_{P}^{2}R}{2}+ \Theta(X) R_{GB}^2\right] +S_{\text{Matter}},
\end{equation}
where
\begin{equation}\label{thetad}
\Theta(X)\equiv f(X)-B\frac{\varphi}{M},
\end{equation}
where $B$ is a dimensionless constant and we have set $\hbar=c=1$.\\
In addition to the shift symmetry for $\varphi$, the action is also invariant under $f\rightarrow f+c$, where $c$ is an arbitrary constant.
The main aim of this paper is to show that the scenario which is described in Fig. 1 is achievable through this model.
At first glance, the condensation in the model may seem irrelevant since it is expected that at the condensed phase $f(X)R_{GB}^2$ has no effect on dynamics of the model. However, we will show that  $\varphi$ can be condensed by using a mechanism which is different than the mechanism which is used in the ghost condensate scenario.\\
As we will see, because of the symmetries, the corresponding energy density of the condensed state will be $\rho_{\varphi}=3M^2M_P^2$. Thus, the energy density of the condensed state dose not depend on $B$ and  parameters which describe of $f(X)$. In other words, at the condensed phase the model has not free parameters  and is the same as $\Lambda CDM$ model.\\  
The organization of this paper is as follows: in Sec. 2 we study background cosmology of the model. Sec. 3 is devoted to study the dynamics of the model with perturbed metrics. Finally, in Sec. 3 we summarize our finding.
\section{Background equations}
As we pointed out, the main goal of this paper is to justify the scenario which is depicted in Fig.1. As the background, we use the flat Friedmann-Robertson-Walker (FRW) metric as
\begin{equation}\label{1-1}
ds^{2}=-dt^{2}+a(t)^{2}dx^{i}dx^{j}\delta_{ij},
\end{equation}
where $a=a(t)$ is the scale factor that from which the Hubble parameter is defined as $H\equiv\dot{a}/a$, where an overdot denotes for a derivative with respect to $t$ .\\
\begin{figure}[!h]
\includegraphics[width=0.5\textwidth]{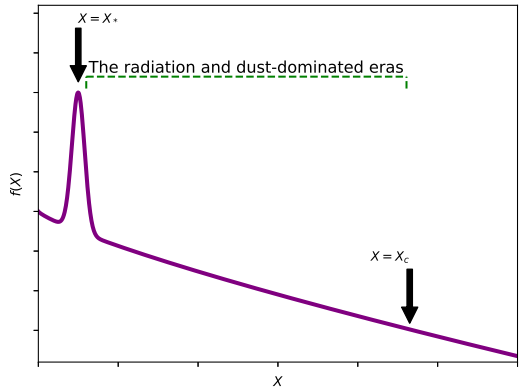}
\caption{Schematic of the model. The figure can be obtained from Eq. \eqref{p} with $n=20$, $\epsilon=0.5$, $\alpha=0.8$. As an initial value, one can take $X\approx X_*$. During the primordial inflation, this point is an attractor solution and the scalar field keeps its value. During the radiation-dominated era, $X=X_*$ changes to an unstable point and the scalar field ``rolls down $f(X)$`` . During the radiation and dust-dominated eras, $\rho_{\varphi} $ decreases at a slower rate than the background density. Eventually, $\rho_{\varphi} $ dominates over the the background density. At $X=X_c$, the scalar field could be condensed with  $\rho_{\varphi}=3M_P^2H_0^2 $ .} 
\end{figure}
The present Hubble parameter is $H_0\approx 10^{-33}eV$. The scale factor for the radiation-dominated era is $a=(Mt)^\frac{1}{2}\ll 1$ and for the dust-dominated era is $a=(Mt)^\frac{2}{3}\ll 1$ such that $a_0\approx1$. Also, in this paper by "late-time" we mean situations for which $a\geq 1.$\\
To describe cosmology for $a>1$, the corresponding energy density of the modified part in the EH action $\rho _\varphi$ must be tracked the energy density of radiation $\rho_r$ and the energy density of dust $\rho_d$ 
in such a way that eventually $\rho_r$ and  $\rho_d$ will fall below  $\rho_{\varphi}$. Hence, during the radiation and dust-dominated eras, all components of $\rho_{\varphi}$ must be gone as $a^{-y}$, where $y\leq 3$.
we will introduce the model in such a way that satisfies the stated conditions.\\
Varying the action \eqref{i-0} with respect to the metric  
and using the following identity in four dimension
\begin{equation}
R_{GB}^2 g_{\mu\nu}=4RR_{\mu\nu}-8R_\mu^\rho R_{\nu\rho}-8R^{\rho\sigma}R_{\rho\mu\sigma\nu}+4R_{\rho\sigma\tau\mu}R^{\rho\sigma\tau}_{\nu},
\end{equation}
yields
\begin{equation}\label{B1}
\begin{split}
R_{\mu\nu}-\frac{1}{2}g_{\mu\nu}R=\frac{1}{M^2_P}&[4R\nabla_\mu\nabla_\nu\Theta-4Rg_{\mu\nu}\nabla^2\Theta-8R^\rho _\nu \nabla_\rho\nabla_\mu \Theta-8R^{\rho} _{\mu} \nabla_\rho\nabla_\nu \Theta\\
&+8R_{\mu\nu}\nabla^{2}\Theta+8g_{\mu\nu}R^{\alpha\beta}\nabla_\alpha\nabla_\beta\Theta+8R_{\mu\;\;\;\nu}^{\;\rho\sigma}\nabla_\rho\nabla_\sigma\Theta]
+\frac{f_X}{M_{P^2}M^{4}}R^2_{GB}\nabla{_\mu}\varphi\nabla{_\nu}\varphi\\
&+\frac{T_{\mu\nu}}{M^2_P},
\end{split}
\end{equation} 
where $f_X$ denotes for derivative with respect to $X$ and $T_{\mu\nu}$ is the corresponding energy-momentum tensor for the radiation or dust.\\
Varying \eqref{i-0} with respect to $\varphi$ and using FRW metric gives
\begin{equation}\label{B2}
(3H\dot{\varphi}+\ddot{\varphi})f_X+2\ddot{\varphi}Xf_{XX}+\dot{\varphi}f_X\frac{\dot{R}^2_{GB}}{R^2_{GB}}=-BM^3,
\end{equation}
where, for FRW metric, the Gauss-Bonnet term takes the following form
\begin{equation}\label{gbq}
R^2_{GB}=24H^2(\dot{H}+H^2).
\end{equation}
If we take $\rho_{\text{Matter}}$ as the energy density of  radiation and dust, the $(t,t)$ component of \eqref{B1} gives 
\begin{equation}\label{B4}
H^2=\frac{\rho_{\text{Matter}}}{3M_{P}^2}+\frac{16H^4}{M_P^{2}}\big[(1+\frac{\dot{H}}{H^2})-\frac{\ddot{\varphi}}{H\dot{\varphi}}\big]Xf_X+\frac{8BH^3\dot{\varphi}}{MM_{P}^2}. 
\end{equation}
The above expression can be used to define the energy density of the scalar field as
\begin{equation}\label{BF}
\rho_\varphi\equiv 48H^4[(1+\frac{\dot{H}}{H^2})-\frac{\ddot{\varphi}}{H\dot{\varphi}}]Xf_X+\frac{24BH^3\dot{\varphi}}{M}.
\end{equation}
Note that if $H\rightarrow 0$ then $\rho_{\varphi}\rightarrow 0$. So, for situations for which the expansion of the universe is suppress and the spacetime can be described by the Minkowski metric, one can take $\rho_{\varphi}=0.$\\   
Let us assume that $f(X)$ has a maximum at $X=X_*$ and, as an initial condition, setting $X|_{\text{initial}}=X_*$ as is shown in Fig. 1. From Eq. \eqref{B2} it follows that very close to $X=X_{*}$ we have
\begin{equation}\label{B3}
\ddot{\varphi}_*\approx\frac{-BM^3}{2X_*f_{XX}|_{X= X_{*}}}.
\end{equation}
Since $f_{XX}|_{X= X_{*}}<0$, for $B>0$ we find $\ddot{\varphi}_*>0$.\\
The stated initial conditions will be used to study the dynamics of the scalar field. As we will show, when the expansion is
dominated by the scalar field, $\rho_\varphi$ dose not depend on details of $f(X)$. This property allow us to choose suitable forms for $f(X)$  . \\
In what follows, we try to obtain general results. However, to investigate the consequences of the equations, we assume that $f(X)$ has the Gaussian distribution  around $X_*$ with a ``\textit{tilted path}`` after $X_*$, as is shown in Fig. 1. A prototypical form can be provided by
\begin{equation}\label{p}
f(X)=exp[-n(X-X_*)^{2}]-\epsilon X^{\alpha}.
\end{equation}
If $\frac{\epsilon\alpha}{n}\ll1$, the maximum of $f(X)$ is given by $X\approx X_*$. Also, for $n>1$ and $X>X_*$, it follows that  $f(X)\approx-\epsilon X^{\alpha}$. To have the tilted path after $X_*$, we demand $\alpha\leq1$ .\\
Using the above consideration, the dynamics of the scalar field is as follows:
\begin{itemize}
\item \emph{the inflationary era}: for the primordial inflation that for which $\epsilon\equiv\frac{|\dot{H}|}{H^2}\ll1, \frac{\dot{\epsilon}}{H\epsilon}\ll1$, by using Eq. \eqref{gbq} it is easy to show that $\frac{\dot{R}^2_{GB}}{R^2_{GB}}\ll H$. Therefor, during the inflation, the last term in left-hand side of Eq. \eqref{B2} can be neglected compared with the first term in left-hand side of Eq. \eqref{B2}. So, we have
\begin{equation}\label{Bi2}
(3H\dot{\varphi}+\ddot{\varphi})f_X+2\ddot{\varphi}Xf_{XX}\approx-BM^3.
\end{equation}
Now, consider a small perturbation around $X_*$ as $X=X_*+\delta X$. Thus, Eq. \eqref{Bi2} results in 
\begin{equation}\label{eqin}
\delta \dot{X}+3H\delta X+MW(X_*)\delta X=0,
\end{equation}
where
\begin{equation}
W(X_*)\equiv 2\frac{\ddot{\varphi_*}}{M\dot{\varphi}_*}\left(1+X_*\frac{f_{XXX}}{f_{XX}}\big|_{X=X_*}\right).
\end{equation}
Eq. \eqref{eqin} has the following solution
\begin{equation}
\delta X=\delta X_* e^{-3 \mathcal{N}}e^{-MW(X_*)\Delta t},
\end{equation}
where $\mathcal{N}$ is the number of $e-$foldings of expansion during inflation and $\Delta t$ is the time interval between the beginning and end of inflation. Also, $\delta X_*$ is the initial value for $ \delta X$.\\
So, if $W(X_*)\geq 0$ , $X_*$  is stable and attractor solution during inflationary era. We will take $W(X_*)\approx 0$, which can be used to impose some conditions on the parameters of $f(X)$ around $X=X_*$. It is clear that $W(X_*)\approx 0$ has to be
considered just as a condition at $X=X_*$ for $f(X)$. For Eq. \eqref{p} this condition satisfies by taking $\frac{\epsilon\alpha}{n}\ll1$.\\ 
Therefore, by imposing the stated condition during inflationary era, the energy density of the scalar field becomes  
\begin{equation}\label{BFcc}
\rho_\varphi|_{\text{inflation}}=\frac{24BH^3\dot{\varphi}_*}{M}.
\end{equation}
To have radiation-dominated era after the inflationary phase, we have to impose the following condition
\begin{equation}\label{condition}
\rho_{\varphi}|_{\text{end of inflation}}\ll\rho_{\text{radiation}}.
\end{equation}
If $H_*$ is the value of Hubble parameter at the end of inflation, then $\rho_{\text{radiation}}\approx H_*^4$. So, Eq. \eqref{condition} gives an upper bound for $\dot{\varphi_*}$ as
\begin{equation}
\dot{\varphi_*}\ll \frac{MH_*}{24B}.
\end{equation} 
\item  \emph{The radiation-dominated era:} Imposing Eq. \eqref{condition} leads us to radiation-dominated universe after inflation that for which $ H=M/(2a^2)$. Considering $X=X_*+\delta X$ and using Eq. \eqref{B2} results in
\begin{equation}\label{Bi3}
\delta\dot{ X}-\frac{5}{2t}\delta X+MW(X_*)\delta X=0,
\end{equation}
which has the following solution
\begin{equation}
\delta X=\delta X_*(\frac{a}{a_*})^{5}e^{-W(X_*)(a^2-a_*^2)}.
\end{equation}
The above result shows that if $W(X_*)\approx 0$, $X= X_*$ changes to an unstable solution. Note that, as follows from Eq. \eqref{B3}, $\ddot{\varphi_*}>0$. So, by imposing continuity condition, $\dot{\varphi}$ and $X$ grow with time, as is shown in Fig. 1.\\
Since $X_*$ is the unstable solution during the radiation-dominated era, linearizion about $X_*$ is not valid and Eq. \eqref{B2} must be solved exactly. Eq. \eqref{B2} takes the following form
\begin{equation}\label{radeq}
\frac{d}{dt}(\dot{\varphi}f_X)-\frac{5}{2}\dot{\varphi}f_X=-BM^3.
\end{equation}
Using the initial conditions, $X|_{t=t_*}=X_*$, the above equation has a solution as
\begin{equation}\label{rads}
\dot{\varphi}f_X=\frac{2}{3}BM^2a^2(1-\frac{a^3}{a^3_*}),
\end{equation}
where $a_*=a|_{t=t_*}$.\\
Combining Eqs. \eqref{BF} and \eqref{rads}, gives
\begin{equation}\label{radrq}
\rho_\varphi=\frac{B\dot{\varphi} M^2}{a^8}\big[a^2(1+\frac{\ddot{\varphi}}{H\dot{\varphi}})(\frac{a^3}{a^3_*}-1)+3\big].
\end{equation}
To obtain $\rho_\varphi$ in terms of $a$, $\dot{\varphi}$ must be obtained from Eq. \eqref{rads}. To solve this equation, note that almost after the beginning of the radiation-dominated era, $(\frac{a}{a_*})^3\gg1$. So, one can drop the term +1 in square parentheses in Eq. \eqref{rads}. Hence, the absolute value of the left-hand side of the above equation is increased as $(\frac{a}{a_*})^5\gg1$. So, $f(X)$ must be chosen in such a way that $\dot{\varphi}f_X$ is increasing. If we use Eq. \eqref{p} with $n>1$, it turns out that soon after the beginning the radiation-dominated era, one can drop the exponential term in Eq. \eqref{p}. Combining   these approximations with Eqs. \eqref{p} and \eqref{rads} results in
\begin{equation}\label{rads1}
\dot{\varphi}\approx M^2\big[\frac{2^{\alpha}B}{3\epsilon\alpha}a_*^2\big]^{\frac{1}{2\alpha-1}}(\frac{a}{a_*})^{\frac{5}{2\alpha-1}}\hspace{0.3cm},
\end{equation}
which gives
\begin{equation}\label{rads2}
\frac{\ddot{\varphi}}{H\dot{\varphi}}=\frac{5}{2\alpha-1}.
\end{equation}
It follows then from  Eqs. \eqref{radrq}, \eqref{rads1} and \eqref{rads2} that
\begin{equation}
\rho_\varphi\approx M^{4}\frac{B^{\frac{2\alpha}{2\alpha-1}}}{a_*^8}
       \Big[\frac{2^{\alpha}a_{*}^2}{3\epsilon\alpha}\Big]^{\frac{1}{2\alpha-1}}\Big[\frac{2\alpha+4}{2\alpha-1}a_*^2(\frac{a_*}{a})^\frac{2(3\alpha-4)}{2\alpha-1}+3(\frac{a_*}{a})^\frac{16\alpha-13}{2\alpha-1}\Big].
\end{equation}
So, if
\begin{equation}\label{rban}
\frac{1}{2}<\alpha\leq1,
\end{equation}
$ \rho_\varphi $ decreases at a slower rate than the background density in the radiation-dominated era.\\
\item \emph{The dust-dominated era}: During dust-dominated era, for which $H=\frac{2}{3t}$, $a\propto t^{2/3}$, $H=\frac{2M}{3a^\frac{3}{2}}$, Eq. \eqref{B2} becomes
\begin{equation}
\frac{d}{dt}(\dot{\varphi}f_X)-\frac{2}{t}f_X=-BM^3,
\end{equation}
which has the following solution
\begin{equation}\label{d3}
\dot{\varphi}f_X=C a^3+BM^2a^{\frac{3}{2}},
\end{equation} 
where $C$ is an arbitrary constant of integration.\\
To determine $C$, we take $a_{\text{eq}}$ as the scale factor at dust-radiation equality, which is  $a_{\text{eq}}\approx10^{-4}$ \citep{pl}. Now, consider $a_{\text{eq}}$ as the beginning of the dust-dominated era. So, at $a=a_{\text{eq}}$, the right-hand side of Eq. \eqref{rads} must be equal to the right-hand side of Eq. \eqref{d3}. This condition yields
\begin{equation}
C=-\frac{BM^2}{a^\frac{3}{2}_{\text{eq}}}\Big[\frac{2a^{\frac{1}{2}}_\text{eq}}{3}(\frac{a_{\text{eq}}}{a_*})^3-1\Big]\approx-\frac{2BM^2}{3a_{\text{eq}}}(\frac{a_{\text{eq}}}{a_*})^3.
\end{equation}
Thus
\begin{equation}\label{d3}
\dot{\varphi}f_X=BM^2\Big[a^{\frac{3}{2}}-\frac{2a^3}{3a_{\text{eq}}}(\frac{a_{\text{eq}}}{a_*})^3 \Big].
\end{equation} 
Even at $a=a_{\text{eq}}$ the first term in the right-hand side of the above equation in less than the other term. So
\begin{equation}
\dot{\varphi}f_X\approx-\frac{2BM^2}{3a_{\text{eq}}}(\frac{a_{\text{eq}}}{a_*})^3 a^3.
\end{equation}
Using the above equation with Eq. \eqref{p}, for $X>X_*$, gives
\begin{equation}\label{duas11}
\dot{\varphi}\approx M^2\big[\frac{2^{\alpha}B}{3\epsilon\alpha}a^2_{\text{eq}}\big]^{\frac{1}{2\alpha-1}}(\frac{a}{a_*})^{\frac{3}{2\alpha-1}}\hspace{0.3cm}.
\end{equation}
Thus
\begin{equation}\label{du111}
\frac{\ddot{\varphi}}{H\dot{\varphi}}=\frac{3}{2\alpha-1}.
\end{equation}
Combining Eqs. \eqref{BF} , \eqref{duas11} and \eqref{du111}, yields
\begin{equation}
\rho_\varphi=\frac{64}{9}M^4\Big[\frac{2^{\alpha}B}{3\epsilon\alpha}a^2_{\text{eq}}\Big]^{\frac{1}{2\alpha-1}}\Big[\frac{2\alpha+5}{4\alpha-2}\hspace{0.2cm}\frac{4}{9a_{\text{eq}}}(\frac{a_{\text{eq}}}{a_*})^2a^{\frac{6-6\alpha}{2\alpha-1}}+a^{\frac{15-18\alpha}{4\alpha-2}}\Big]. 
\end{equation}
So, if
\begin{equation}\label{dban}
\frac{1}{2}<\alpha\leq\frac{3}{2},
\end{equation}
$\rho_\varphi$ decreases at a slower rate than the background density.\\
Comparing Eq. \eqref{rban} with Eq. \eqref{dban}, results in
 \begin{equation}\label{fban}
\frac{1}{2}<\alpha\leq 1.
\end{equation}
Therefore, by taking the above result for Eq. \eqref{p}, eventually $\rho_\varphi$ dominates over the radiation and dust densities. 
\item \emph{The condensed phase}: by taking $\rho_{\text{Matter}}\approx 0$ in Eq. \eqref{B4}, we have
\begin{equation}\label{Bc4}
H^2\approx\frac{16H^4}{M_P^{2}}\big[(1+\frac{\dot{H}}{H^2})-\frac{\ddot{\varphi}}{H\dot{\varphi}}\big]Xf_X+\frac{8BH^3\dot{\varphi}}{MM_{P}^2}. 
\end{equation}
Note that if $\dot{\varphi}\to 0$ then $H\to 0$. Also, if $B=0$ and taking $f_X\to 0$ it follows that $H\to 0$. we will not consider such situations.\\
Dividing Eq. \eqref{Bc4} by $H^2$ yields
\begin{equation}\label{Bc44}
\frac{16H^2}{M_P^{2}}\big[(1+\frac{\dot{H}}{H^2})-\frac{\ddot{\varphi}}{H\dot{\varphi}}\big]Xf_X+\frac{8BH\dot{\varphi}}{MM_{P}^2}\approx 1. 
\end{equation}
Before we study the above equations by a systematic method, let us seek for a solution by taking $\ddot{\varphi}=0$, $\dot{H}=0$, but with $\dot{\varphi}=\dot{\varphi}_{c}\neq 0$. From Eq. \eqref{B2} and our demands it follows that
\begin{equation}\label{Bc5}
3H_c\dot{\varphi}_cf_X(X_c)=-BM^3,
\end{equation}
where subscript $c$ stands for ``condensed``.\\ 
From the above conditions and Eq. \eqref{Bc44} we have
\begin{equation}\label{Bc55}
\dot{\varphi}_c=\frac{3MM_{P}^2}{16BH_c}.
\end{equation} 
Note that the relation between $\dot{\varphi}_c$ and $H_c$ does not depend on $f(X)$.\\
Using the definition of $\Theta$ , the above  condition can be rewritten as
\begin{equation}\label{Bc6}
\frac{8H_c\dot{\Theta}_c}{M_P ^2}=\frac{-3}{2},
\end{equation}
Eqs. \eqref{Bc5} and \eqref{Bc55} results in
\begin{equation}\label{Bc7}
f_X(X_c)=-\frac{16}{9}B^2\frac{M^2}{M_{P} ^2}.
\end{equation}
From Eqs. \eqref{BF}, \eqref{Bc55} , \eqref{Bc7} it follows that
\begin{equation}
\rho_{\varphi}=3H_{c}^2M_P^2.
\end{equation}
Obviously, $\rho_{\varphi}$ is constant and dose not depend on $f(X)$ or $B$. If we take $H_c=H_0\approx M$,  then $\rho_{\varphi}=\rho_{\Lambda}$.\\
As is clear from  Eqs. \eqref{Bc5} and \eqref{Bc6},    
to have the above result, it is necessary that $H_c\neq 0$ which shows the difference between the $\Lambda$CDM and this model. We pointed out that when $H\rightarrow 0$, then $\rho_{\varphi}\rightarrow 0$. 
\end{itemize}
\subsection{Systematic study of equations}
The condensed state is not a unique solution for the background equations. To use it for the late-Universe, the condensed state must be an attractor solution. To investigate this issue, let us introduce $\tau=Mt$ as the dimensionless time with the following dimensionless variables 
\begin{equation}
x\equiv\frac{\dot{\varphi}}{M^2},\hspace{0.75cm} y\equiv\frac{H}{M},\hspace{0.75cm} z\equiv y', 
\end{equation}
where prime denotes the derivative with respect to $\tau$.\\
Using the above variables, Eqs. \eqref{B2} and \eqref{Bc44} can be written in the following autonomous form
\begin{equation}\label{sys1}
  \begin{cases} 
   x'=g_1 &  \\
   y'=g_2      & \\
   z'=g_3
  \end{cases},
\end{equation}
where
\begin{equation}
g_1=xy+\frac{zx}{y}+\frac{B}{f_X}-\frac{M_P^2}{M^2}\frac{1}{8xyf_X},
\end{equation}
\begin{equation}
g_2=z,
\end{equation}
\begin{equation}
g_3=-\left(\frac{B}{xf_X}+3y+\frac{g_1(x,y,z)}{x}+xg_1(x,y,z)\frac{f_XX}{f_X}+\frac{2z}{y}\right)(z+y^2)-2yz.
\end{equation}
To find the fixed points, we have to solve $x'=0, y'=0, z'=0$. These equations give us a nontrivial fixed point as
\begin{equation}\label{sys2}
3y_c=-\frac{B}{x_Cf_X},\hspace{0.75cm} Bx_Cy_c=\frac{3}{16}\frac{M_P^2}{M^2},
\end{equation}
which are Eqs. \eqref{Bc5} and \eqref{Bc55} in terms of the dimensionless variables. Thus, the nontrivial fixed point describes the condensed state.\\
To study the stability of the fixed point, consider small perturbations  around the fixed point as $x=x_c+\delta x, y=y_c+\delta y, z=z_c+\delta z$, and then substituting them into Eq. \eqref{sys1}. The stated procedure and using Eq. \eqref{sys2} yields
\begin{equation}
\frac{d}{d\tau} \begin{pmatrix}
  \delta x  \\
  \delta y \\
  \delta z
 \end{pmatrix}= T_c
 \begin{pmatrix}
  \delta x \\
  \delta y \\
 \delta z
 \end{pmatrix},
\end{equation}
where
\begin{equation}
T_c = 
 \begin{pmatrix}
  y_c(-1+q) & -x_c  & \frac{x_c}{ y_c} \\
 0 & 0 &  1\\
  \frac{\partial g_3}{\partial x}|_{x=x_c,y=y_c,z=z_c} & -2+\frac{x_cq}{2}\hspace{0.5cm} & \hspace{0.5cm}-5y_c-qy_c
 \end{pmatrix},
\end{equation}
where
\begin{equation}
q\equiv\frac{2Xf_{XX}}{f_X}|_{X=X_c}.
\end{equation}
The stability of the fixed point is determined by the eigenvalues of $T_c$ . If all eigenvalues have a negative real part, the fixed point is the attractor solution.
The eigenvalues of $T_c$, $\lambda$, will satisfy the following equation 
\begin{equation}
\lambda^3+6y_c\lambda^2+\lambda[7y_c^2+2-q(1-y_c^2)]+2y_c(1-y_c^2)+qy_c[(q-3)-(q+3)y_c^2]=0.
\end{equation}
If we set $H_c=M$, which means $y_c=1$, it follows that
\begin{equation}\label{wer}
\lambda(\lambda+3)^2=6q.
\end{equation}
Since $X_c\gg X_*$, from Eqs. \eqref{p} and \eqref{wer} we find
\begin{equation}\label{rootb}
\lambda(\lambda+3)^2=12(\alpha-2).
\end{equation}
For $0.5<\alpha\leq 1$, Eq. \eqref{rootb} has solutions with a negative real part. For example, by taking $\alpha=0.75$, the above equation yields  
\begin{equation}
\lambda_1\approx -4.77,\hspace{0.5cm}\lambda_2\approx-0.61-1.66I,\hspace{0.5cm}\lambda_3\approx -0.61+1.66I.
\end{equation}
Table I shows solutions of Eq. \eqref{rootb} for some values of $\alpha$ in $0.5<\alpha\leq 1$.\\
\begin{table}[htb]
    \caption{Some solutions of Eq. \eqref{rootb}.}
\begin{tabular}{ |p{0.75cm}|p{1cm}|p{2cm}|p{2cm}|  }
 \hline
 $\alpha$& $\lambda_1$ &$\lambda_2$&$\lambda_3$\\
 \hline
 0.6   & -4.86    &-0.57-1.77$I$& -0.57+1.77$I$ \\
 0.7&   -4.80 & -0.59-1.70$I$  &-0.59+1.70$I$\\
 0.8 &-4.74 & -0.63-1.62$I$&  -0.63+1.62$I$\\
 0.9    &-4.68 &  -0.66-1.54$I$& -0.66+1.54$I$ \\
 1.0& -4.61    & -0.69-1.46$I$&  -0.69+1.46$I$\\
 \hline
\end{tabular}
\end{table}

\section{\label{sec:level1}  Cosmological perturbations}
In this section, we study cosmological perturbations of the model at the condensed phase of the scalar field.
The scale of energy, $\rho_{\varphi}\approx 10^{-3}eV^4$, and the fact that we deal with the de Sitter space impose restrictions on usage of some of the notions. Let us clarify some issues in this context.\\  
For $a\geq 1$, our goal is not study of quantum fluctuations of the perturbed quantities. Also, in contrast with the radiation and dust-dominated eras, in the de Sitter   space any perturbation which has a wavelength greater than the Hubble horizon will never enter the horizon. Therefore, it is sufficient to consider perturbations which are inside the horizon. Although these perturbations will eventually exit from the Hubble horizon, and the universe is going to the pure de Sitter space, but study of them will provide  information about the model.\\
Moreover, at late-times, there exist a cosmological scale, $r_h\approx100Mpc$, for which the FRW metric can be applied as the background metric. For $r\ll r_h$, the spacetime is flat \footnote{Except very close to a black hole which lies out of the scope of this paper. }.\\
Note that, in the previous section it has been shown that  by taking $t_0\approx 1/H_c$ as the age of the universe, for $t>t_0$ the condensed phase could be formed. So, the perturbed equations in this section must be considered for $t\geq t_0$.\\
As we will see, the perturbed equations have the following form
\begin{equation}\label{modc1}
\delta G_{\mu\nu}=\frac{8H_c\dot{\Theta}_c}{M_P^2}\delta Y_{\mu\nu},
\end{equation}
where $\delta G_{\mu\nu} $ is the perturbed Einstein tensor and $\delta Y_{\mu\nu}$ is the corresponding perturbed expression in each sector( see below). From Eq. \eqref{Bc6} it is clear that $\frac{8H_c\dot{\Theta}_c}{M_P^2}$ is just a number and this may lead to a miss interpretation of the results
. To clarify the issue, note that $H_c$ at the right-hand side of Eq. \eqref{modc1} shows that the right-hand side of this expression vanishes for $r\ll r_h$ that for which the  gravitational attraction binds matter together. As we have mentioned, for $t>t_0$ and $r\ll r_h$ the spacetime is flat.\\
To consider the above circumstances, Eq. \eqref{modc1} can be represented in two ways as follows :
\begin{itemize}
\item the first way is similar to Ref. \citep{Arkani}. For this purpose, note that one can take $H_c=M$, and $\dot{\Theta}_c=M\Theta_c'$. Now, Let us define the following variables
\begin{equation}
\Gamma_{H_c}\equiv\frac{H_c M^2}{M_P^2} ,\hspace{2cm}m\equiv\frac{M^2}{\sqrt{2}M_P} .
\end{equation}
$\Gamma_{H_c}^{-1} $ can be regarded as`` timescale over which modifications of gravity
take place, at a length scale of order $m^{-1}$`` \citep{Arkani}.
Thus
\begin{equation}\label{modc2}
\delta G_{\mu\nu}=\frac{4\Gamma_{H_c}^2}{m^2}\Theta_c'\delta Y_{\mu\nu}.
\end{equation}
\item The other way is to rewrite Eq. \eqref{modc1} as
\begin{equation}
   \left\{
   \begin{array}{cc}
     \delta G_{\mu\nu}\approx0 & r < r_h\hspace{.15cm}(t\geq t_0)  \\
     \delta G_{\mu\nu}=\frac{8H_c\dot{\Theta}_c}{M_P^2}\delta Y_{\mu\nu} &  r \geq r_h \hspace{.15cm}(t\geq t_0)
   \end{array}
   \right.
\end{equation}
\end{itemize}     
\subsection{\label{sec:level1}  The scalar metric perturbations}
For the scalar metric perturbations, it is convenient to use the Newtonian gauge at late-times \citep{pl}. This gauge is defined as
\begin{equation}
ds^{2}=-(1+2\Phi^N(t,x))dt^{2}+a(t)^{2}(1-2\Psi^N(t,x))dx^{i}dx^{j}\delta_{ij}.
\end{equation}
where $N$ stands for ``Newtonian``.\\
A measurable quantity for this case is 
\begin{equation}
\gamma=\frac{\Psi^N}{\Phi^N},
\end{equation}
that for which $\Lambda$CDM model predicts that $\Psi^N=\Phi^N$.
From the Planck CMB temperature data, the current value for $\gamma$ reported as \citep{pl}
\begin{equation}\label{plp}
\gamma_0-1=0.70\pm 0.94.
\end{equation}
Also, if we consider the weak lensing data we have \citep{pl}
\begin{equation}
\gamma_0-1=1.36^{+1.0}_{-0.69}.
\end{equation}
To find $\gamma$, we first use the comoving gauge for which $\delta\varphi=0$ and can be parameterized as
\begin{equation}\label{gco}
ds^{2}=-(1+2\Phi(t,x))dt^{2}+2a(t)\partial_{i}K(x,t)dtdx^i+a(t)^{2}(1+2\zeta(t,x))dx^{i}dx^{j}\delta_{ij}.
\end{equation}
Then we get from the comoving gauge to the Newtonian gauge by
\begin{equation}\label{trg}
\Phi^N(t,x)=\Phi(x,t)+\frac{d}{dt}(aK(x,t)),\hspace{1cm}\Psi^N(t,x)=-\zeta(t,x)-HaK(x,t).
\end{equation}
Furthermore, the Fourier components of a general perturbation $Q(x,t)$  is defined as
\begin{equation}
Q=\int Q(t,x)e^{-I\textbf{k.x}}d^3x.
\end{equation}
With the substitutions Eq. \eqref{gco} and taking $\mu=0,\nu=i$ in Eq. \eqref{B1} we obtain
\begin{equation}
\delta G_{0i}=\frac{-8H_c\dot{\Theta}_c}{M_P^2}\Big[\delta G_{0i}+H_c^2 a\partial_{i}K(x,t)+H_c\partial_i\Phi(t,x)\Big].
\end{equation}
The above equation and Eq. \eqref{Bc6}, in terms of the Fourier components, give
\begin{equation}\label{pra}
H\Phi=\frac{2}{5}\dot{\zeta}. \hspace{2 cm}\text{ for $r\geq r_h,t\geq t_0$}
\end{equation}
Eq. \eqref{gco} and taking $\mu=i\neq \nu=j$ in Eq. \eqref{B1} results in
\begin{equation}\label{perttw}
\delta G_{ij}=\frac{8H_c\dot{\Theta}_c}{M_P^2}\Big[\partial_i\partial_j\Phi(x,t)+a\partial_i\partial_j \dot{K}(x,t)+2H_ca\partial_i\partial_jK(x,t)\Big].
\end{equation}
Using Eqs. \eqref{Bc6} and \eqref{perttw}, in terms of the Fourier components, it turns out that
\begin{equation}\label{pij}
\frac{\Phi}{2}+H_c a K+\frac{a\dot{K}}{2}=\zeta. \hspace{2 cm}\text{ for $r\geq r_h,t\geq t_0$}.
\end{equation}
Finally, by taking $\mu=\nu=i$ in Eq. \eqref{B1} we have
\begin{equation}\label{pii}
\begin{split}
\delta G_{ii}=\frac{8H_c\dot{\Theta}_c}{M_P^2}&\Big[a^2\delta R+5H^2\delta g_{ii}-\delta R_{ii}+10a^2H^2\Phi(x,t)-12a^2H\dot{\zeta}(x,t)+3Ha\partial^2K(x,t)\\
&+\partial^2\Phi(x,t)+3\partial^2\zeta(x,t)+\partial^2\dot{K}(x,t)+2Ha^2\dot{\Phi}(x,t)-3\ddot{\zeta}(x,t)-\partial_i\partial_i\zeta(x,t)\Big],
\end{split}
\end{equation}
where $\partial^2\equiv\partial_i\partial_i$.\\
Combining Eqs. \eqref{Bc6}, \eqref{pij} and \eqref{pii}, and then using the Fourier components of the perturbed quantities, gives
\begin{equation}\label{pre}
2\ddot{\zeta}+6H\dot{\zeta}-12H^2\Phi-5H\dot{\Phi}=0. \hspace{2 cm}\text{ for $r\geq r_h,t\geq t_0$}
\end{equation}
From Eqs. \eqref{pra} and \eqref{pre} it follows that
\begin{equation}
\dot{\zeta}=0\rightarrow \zeta=\zeta_0. \hspace{2 cm}\text{ for $r\geq r_h,t\geq t_0$}
\end{equation}
Now, from the above results, one can find $\Phi$ and $K$ as 
\begin{equation}
\Phi=0,\hspace{0.4cm} aK=C_{2} e^{-H_c t}+\frac{2\zeta_0}{H_c},\hspace{2 cm}\text{ for $r\geq r_h,t\geq t_0$}
\end{equation}
where $C_2$ is an arbitrary constant.\\
So, by using Eq. \eqref{trg}, it follows that
\begin{equation}
\Phi^N=-H_c C_2 e^{-H_c t},\hspace{.4cm}\Psi^N=-3\zeta_0+\Phi^N.  \hspace{2 cm}\text{ for $r\geq r_h,t\geq t_0$}
\end{equation}
Therefore, the model predicts that
\begin{equation}
\gamma-1=\frac{3\zeta_0}{H_c C_2} e^{H_c t}.  \hspace{2 cm}\text{ for $r\geq r_h,t\geq t_0$}.
\end{equation}
Since $H_c t_0\approx 1$, From Eq. \eqref{plp}, we have
\begin{equation}
\frac{3\zeta_0}{H_c C_2}\approx 0.7\pm 0.94.
\end{equation}
Since the anisotropic stress, $\Sigma$, is defined as
\begin{equation}
\Psi^N-\Phi^N=\frac{a^2}{M_P^2}\delta\Sigma,
\end{equation}
the model predicts that if $\zeta_0\neq 0$, we have the induced anisotropic stress in the model.
\subsection{\label{sec:level1}  The tensor metric perturbations}
The tensor metric perturbations,$\gamma_{ij}$, are characterized by
\begin{equation}\label{defmett}
ds^{2}=-dt^{2}+a^{2}[\delta_{ij}+\gamma_{ij}(x,t)]dx^{i}dx^{j},
\end{equation}
where $\partial_{i}\gamma_{ij}=\gamma_i^i=0$.\\
Using polarization states $ e_{i,j}^{(+,\times)}$ as
\begin{equation}\label{tenct}
\gamma_{ij}(x,t)=\int \frac{d^3k}{(2\pi)^{3/2}}\sum_{s=+,\times} e^s_{i,j}\gamma e^{I\textbf{k.x}},
\end{equation}
Eqs. \eqref{B1}, \eqref{defmett} and \eqref{tenct} give
\begin{equation}
\delta G^i_j=\frac{-8H_c\dot{\Theta_c}}{M_P^2}\Big[\ddot{\gamma}+3H_c\dot{\gamma}\Big]=\frac{-4\Gamma_{H_{c}}}{m^2}\Big[\ddot{\gamma}+3H_c\dot{\gamma}\Big].
\end{equation}
Thus
\begin{equation}\label{tc1}
\ddot{\gamma}+3H_c\dot{\gamma}-2\frac{k^2}{a^2}\gamma=0. \hspace{2 cm}\text{ for $r\geq r_h,t\geq t_0$}
\end{equation}
Eq. \eqref{tc1} has the following solution
\begin{equation}\label{ten2}
\gamma =u\hspace{0.12cm}e^{-\frac{3}{2}H_c (t-t_0)},
\end{equation}
where $u$ satisfies the following equation
\begin{equation}\label{ten3}
\ddot{u}-\frac{9H_{c}^{2}}{4}(\frac{8k^2}{9a^2H^2_c}+1)u=0.
\end{equation}
The minus sign of the second term in Eq. \eqref{ten3} shows that if we considered the model for the primordial inflation, we would confront with the instabilities to study gravitational waves \citep{grim}.\\
To investigate the gravitational waves for $a\geq 1$, consider a source of gravitational waves that emits a signal at $t=t_0$. Note that for $r\ll r_h$ (around the source or an observer), we have $ \delta G^i_j=0$ and gravitational waves propagate as usual. Now, let us consider three situations as follows:
\begin{itemize}
\item the observer is located at very far from the source such that $H_c(t-t_0)\gg 1$, where here $t$ is the time that the observer detects the signal.  Since $a=exp[H_c(t-t_0)]$, the redshift will bring each mode outside the horizon as $\frac{k}{a}\rightarrow 0$. So, Eqs. \eqref{ten2} and \eqref{ten3} give
\begin{equation}
\gamma =C_{1T}\hspace{0.15cm}e^{-3H_c(t-t_0)}+C_{2T},
\end{equation}
where $C_{1T}$ and $C_{2T}$ are arbitrary constants.Therefore, outside the horizon the solution becomes a constant.
\item If the distance from the observer to the source is less than $r_h$, the gravitational waves propagate as usual.
\item The modification takes place between the two stated cases. Generally, we have to solve the following equation 
\begin{equation}\label{f}
  \begin{cases} 
   \delta G^i_j=0 & \text{if } r|_{\text{from the source} }\ll r_h \\
   \ddot{\gamma}+3H_c\dot{\gamma}-2\frac{k^2}{a^2}\gamma=0. \hspace{2 cm}&\text{ if $r\geq r_h$}\\
   \delta G^i_j=0&\text{if } r|_{\text{from the observer} }\ll r_h
  \end{cases}
\end{equation} 
The above equation shows that the amplitude of the gravitation wave can be changed. To clarify the issue,  take $\delta t$ as the time interval in which the gravitational wave propagates in $r\geq r_h$. Also, consider    $t-t_0\equiv\delta t<t_0$.  Note that $t_0\approx 10^{10}$year and $H_ct_0\approx1$. For this case   
 $a=exp[{\delta t}/{t_0}]\approx 1$. So, for $r\geq r_h$, we have
 \begin{equation}
 \gamma=C_{3T}\hspace{0.15cm}e^{\frac{-3}{2}\frac{\delta t}{t_0}\Big(1-\frac{\sqrt{8}k}{3aH_c}\Big)}+C_{4T}\hspace{0.15cm}e^{\frac{-3}{2}\frac{\delta t}{t_0}\Big(1+\frac{\sqrt{8}k}{3aH_c}\Big)}.
 \end{equation}
If we take the observed amplitude as $C_{oT}$ and the emitted amplitude as $C_{eT}$ , Eq. \eqref{f} gives relations between the amplitudes. Note that, it is not possible to use this interpretation for the primordial inflation that for which $r_h=0$.
\end{itemize}
\section{\label{sec:level1} Summary}
We have studied a model in which the scalar filed could be condensed and results in a constant value for the Hubble parameter in the late-Universe. During the radiation and dust-dominated eras, the energy density of the scalar field tracks the radiation-dust energy density. The condensed state produces an accelerating universe. The energy density of the scalar field, at the condensed state, is the same as the energy density of the cosmological constant in $\Lambda$CDM model. Moreover, it has been shown that the condensed state is the attractor solution of the equations. We have investigated the cosmological perturbations for the condensed state to seek predictions of the model. It turns out that by measuring the induced anisotropic stress and study evolution of the amplitude of the gravitational waves, one can test the predictions of the model.
\section*{Acknowledgement}
I thank H. Asgari and L. Hu for many useful discussions.

\end{document}